\documentclass[letterpaper, 10pt, conference]{ieeeconf}

\IEEEoverridecommandlockouts
\bstctlcite{BSTcontrol}

\usepackage{graphicx}
\usepackage{amsmath}
\usepackage{url}
\usepackage{booktabs}
\usepackage{makecell }
\usepackage{subcaption}
\usepackage{pifont}

\usepackage{hyperref}
\usepackage{tikz}

\title{\LARGE \bf
  Collecting big behavioral data for measuring behavior against obesity
}

\author{Vasileios Papapanagiotou$^{1}$, Ioannis Sarafis$^{1}$, Christos
  Diou$^{1}$\\Ioannis Ioakimidis$^{2}$, Evangelia Charmandari$^{3}$, Anastasios
  Delopoulos$^{1}$%
  \thanks{*The work leading to these results has received funding from the
    European Community's Health, demographic change and well-being Programme
    under Grant Agreement No. 727688, 01/12/2016 - 30/11/2020
    (http://bigoprogram.eu/).}%
  \thanks{$^{1}$Multimedia Understanding Group, Department of Electrical and
    Computer Engineering, Aristotle University of Thessaloniki, Greece}%
  \thanks{$^{2}$Innovative Use of Mobile Phones to Promote Physical Activity and
    Nutrition Across the Lifespan (the IMPACT) Research Group, Department of
    Biosciences and Nutrition, Karolinska Institutet, 14152 Stockholm, Sweden}%
  \thanks{$^{3}$Division of Endocrinology and Metabolism, Biomedical Research
    Foundation of the Academy of Athens, Athens, 11527, Greece}%
}

\newcommand\copyrighttext{%
  \footnotesize \textcopyright 2020 IEEE.  Personal use of this material
  is permitted. Permission from IEEE must be obtained for all other uses, in
  any current or future media, including reprinting/republishing this material
  for advertising or promotional purposes, creating new collective works, for
  resale or redistribution to servers or lists, or reuse of any copyrighted
  component of this work in other works}
\newcommand\copyrightnotice{%
  \begin{tikzpicture}[remember picture,overlay]
    \node[anchor=south,yshift=10pt] at (current page.south) {\fbox{\parbox{\dimexpr\textwidth-\fboxsep-\fboxrule\relax}{\copyrighttext}}};
  \end{tikzpicture}%
  }

\begin{document}

\maketitle
\thispagestyle{empty}
\pagestyle{empty}
\copyrightnotice

\begin{abstract}
  Obesity is currently affecting very large portions of the global
  population. Effective prevention and treatment starts at the early age and
  requires objective knowledge of population-level behavior on the
  region/neighborhood scale. To this end, we present a system for extracting and
  collecting behavioral information on the individual-level objectively and
  automatically. The behavioral information is related to physical activity,
  types of visited places, and transportation mode used between them. The system
  employs indicator-extraction algorithms from the literature which we evaluate
  on publicly available datasets. The system has been developed and integrated
  in the context of the EU-funded BigO project that aims at preventing obesity
  in young populations.
\end{abstract}

\section{Introduction}
\label{sec:introduction}

Obesity is affecting a very large portion of the population world-wide,
including children and teenagers. According to the World Health Organization,
over $340$ million children were obese or overweight in 2016. Many ``blanket policies'' have been applied so far, but success
rates are low while relapse is quite high. To effectively treat and also prevent
the prevalence of obesity, detailed behavioral profiles on the population level
are needed \cite{bluher2019}. Creating such profiles can answer questions such
as ``what is the average physical activity level of a specific age group in a
specific region'' or ``does visiting fast-food shops and similar food outlets
relate to low physical activity level or obesity''.

Creating population profiles requires collecting large volumes of
individual-level behavioral information. Traditional ways of acquiring such
information are questionnaires, time-use surveys, etc \cite{shim2014dietary},
however, their accuracy has been long challenged
\cite{schatzkin2003comparison}.
On the other hand, technological methods such as the combination of mobile
phones and smart watches that capture wearable signals with relevant
signal-processing and machine-learning algorithms can be a better replacement
\cite{diou2019methodology}. These methods offer many advantages over traditional
surveys. From the participant's view, they reduce the need for user feedback and
also eliminate any personal bias/subjectiveness. From the surveyor's view, they
reduce the effort for gathering the data as well as the need for data curation
(since the gathered data are completely structured). Finally, there are
additional advantages: these methods and tools can be applied to larger
populations, and for longer durations, and are thus able to gather much bigger
volumes of data (big data).

In this work we present a system for collecting big behavioral data from large
populations in order to facilitate large-scale analysis of behavior related to
the prevalence and prevention of obesity. This system has been designed,
implemented, evaluated, and integrated within the context of the EU-funded BigO
project \cite{diou2019methodology}.
It aims at extracting behavior
related to physical activity, types of places that are visited (e.g. parks,
gyms, or fast-foods), and how these places are accessed (i.e. transportation
mode). We propose a set of basic behavioral indicators on the individual level
which can then be used to build population-level behavioral profiles. We
implement and evaluate algorithms from the literature that extract these
individual-level indicators and evaluate them on publicly available datasets in
order to demonstrate the feasibility of such a system.

\section{Extracting behavioral indicators}

Our system aims at collecting data in three main areas of behavior, specifically
physical activity, type of visited places, and transportation mode. Each one of
these has a strong connection to obesity or obesogenic behavior. Particularly,
low level of physical activity is a well established risk factor of obesity
\cite{hills2007, hills2010}.
Thus, we target at estimating the number of steps walked, as well
as the type of physical activity that is performed by the individual.

Knowing what kind of places one visits is also targeted by our system, as it can
be correlated with healthy and non-healthy behavior. For example, there is
evidence of connection between fast-food consumption and obesity
\cite{rosenheck2008fast} that needs to be further explored, while there is
evidence suggesting that availability of certain park facilities plays an
important role in promoting physical activity and healthy weight status
\cite{potwarka2008places}. Additionally, behavior within parks and recreational
facilities can steer design decisions for new parks or facilities to increase
physical activity levels of visitors \cite{besenyi2013observation}.
Thus, we aim at accurately detecting the locations that an individual visits,
based on the location data from his/her mobile phone. The co-ordinates of the
visited locations can then be cross-checked using publicly available map
repositories to derive the type of facility that was visited, i.e. fast-food,
restaurant, park, gym, etc.

In addition to visited locations, our system also aims at detecting the
transportation mode used when individuals move between locations, since evidence
exists that the way one commutes is correlated with overweight and obesity
\cite{lindstrom2008}.

We have selected algorithms for behavioral indicator extraction based on their
reported effectiveness, the clarity of the
relevant publications, and the
results from small-scale comparisons with other counterparts from literature.

\subsection{Physical activity}

For physical activity we focus on counting steps and recognizing the physical
activity type performed at each time and also implement our own algorithm for
activity counts.

The step counting algorithm is presented in \cite{gu2017robust}. It detects
local maxima on the acceleration magnitude, as measured by a mobile phone's 
accelerometer. The local maxima are then filtered against three criteria; the
maxima that satisfy all three criteria are counted are steps.

The first criterion is periodicity: steps are required to occur at a rate
approximately in the range of $1$ to $3$ Hz. The second criterion is similarity:
subsequent steps should exhibit approximately the same level of
acceleration. The algorithm takes into account the steps corresponding each foot
separately, and therefore requires similar acceleration magnitude for the
$1$-st, $3$-rd, $5$-th and so forth steps separately from the $2$-nd, $4$-th,
$6$-th, and so forth steps. Finally, the third criterion is continuity, which
requires that steps occur in groups, and not on their own. We have selected a
threshold of at least $8$ steps in order to count them.

While smart watches are also equipped with  accelerometers, the signals that
they capture are different from the mobile phone. In particular, we can take
advantage of the fixed location of the smart watch and use specialized
step-counting algorithms. The algorithm we have chosen is presented in
\cite{genovese2017smartwatch} and uses a delayed and filtered replica of the
acceleration magnitude in order to extract ``armed'' segments. Each such segment
contains a local maximum, which is counted as a step if it satisfies a series of
pre-defined thresholds.

The physical activity recognition algorithm we use is presented
in \cite{reiss2012creating}. It also operates on the acceleration magnitude of a
 accelerometer: it extracts short, overlapping windows ($5$ s length and $1$ s
step) and estimates several time-domain (such as mean, standard deviation,
peak-to-peak distance, and signal-magnitude area) and frequency-domain (such as
power spectral density) features. The features are then standardized and used in
a multi-class (one-vs-one) SVM classifier in order to infer a single physical
activity class per window. We also apply a majority voting filter of $1$ min
length on the predicted classes to account for some misclassifications.

\subsection{Visited-locations detection}

To detect visited locations from a set of location (GPS) co-ordinates we follow
the approach of \cite{luo2017improved}. The algorithm introduces a metric called
move-ability which is the ratio of actual distance someone moved over the total
traveled distance. Move-ability is then used to compute a density value which is
used with a modified version of the DBSCAN algorithm. By applying the modified
DBSCAN we obtain a set of clusters which correspond to actual visited
locations. The original data (GPS co-ordinates) can then be cross-referenced
against the detected locations in order to obtain the arrival and departure
time-stamps from each location.

\subsection{Transportation-mode recognition}

Recognizing transportation mode is done based on the algorithm of
\cite{shafique2016travel}. However, we first need to detect the trips. We
consider valid trips all the segments from one visited location to the next
(departure from the first until arrival to the next) during which we have no
missing accelerometer data and the average location-speed is at least $0.5$
km/h.

The transportation-mode recognition algorithm of \cite{shafique2016travel}
extracts overlapping windows ($1$ min length and $10$ s step) of the
acceleration magnitude and computes time-domain and frequency-domain features
(similarly to \cite{reiss2012creating}). We have enhanced the set of features
with power spectral density features. We also use multi-class (one-vs-one) SVM
classifier instead of the random forest classifiers used in
\cite{shafique2016travel}. Finally, we perform majority-voting on the detected
transportation mode labels.

\section{Evaluation datasets}

To evaluate the selected algorithms we use three different datasets. The first
one is used for step-counting validation and is published as part of
\cite{brajdic2013walk}. It contains time-annotated sensor signals obtained from
smart phones in typical, unconstrained use while walking. In total, $27$
participants were asked to walk a route at three different walk paces (normal,
fast, and slow). Each participant walked the same distance and changed her/his
speed at markers installed on the path and carried one or two phones placed at
varying positions (in a front or back trouser pocket, in a backpack/handbag, or
in a hand with or without simultaneous typing). This dataset has been used for a
fair, quantitative comparison of standard algorithms for walk detection and step
counting.

The second dataset is the PAMAP2 activity type detection dataset
\cite{reiss2012creating} that contains $12$ activities.  Activity types were
selected to include basic activities (walking, running, traversing stairs),
postures (lying, sitting and standing), common household activities (ironing,
vacuum cleaning), and fitness activities (rope jumping). Each of the subjects
had to follow this protocol, performing all defined activities in the way most
suitable for the subject. Most of the activities from the protocol were
performed over approximately $3$ minutes, except ascending/descending stairs and
rope jumping. Nearly $8$ hours of data were collected altogether.

The dataset used for validating points-of-interest and transportation mode
algorithms is the Sussex-Huawei locomotion (SHL) dataset
\cite{wang2018summary}.
The SHL dataset was collected by the Wearable Technologies Lab at the University
of Sussex as part of a research project funded by Huawei. It is a versatile
annotated dataset of modes of locomotion and transportation of mobile users. It
includes recordings by 3 participants over 3 days in 2017, engaging in 8
different modes of transportation in real-life setting in the United Kingdom.

\section{Preliminary evaluation}
\label{sec:evaluation}

This section presents the evaluation results for the different individual-level
behavioral indicator extraction algorithms on the selected datasets. Table
\ref{tab:step_counting} presents the evaluation of the step-counting algorithm
of \cite{gu2017robust} on the dataset of \cite{brajdic2013walk}. We evaluate
the algorithm on all of the available positions of the mobile phone. Out of all
the different positions, hand-held (with and without using the mobile phone) and
placed in a backpack are the ones with the most available recordings; the error
is quite low in these cases, with less than $10$ steps of absolute error on
average. The handbag placement yields the highest error which can be attributed
to the decreased ``sensitivity'' of the mobile phone's accelerometer when placed
in a bag full of other things and hanged from the subject's shoulder.

\begin{table}
  \centering
  \caption{Step counting results of \cite{gu2017robust} on
    \cite{brajdic2013walk} (mean$\pm$standard deviation).}
  \label{tab:step_counting}
  \begin{tabular}{lrcccc}
    \toprule
    \textbf{Position}
    & \textbf{\#}
    & \makecell{\textbf{Predicted}\\\textbf{steps}}
    & \makecell{\textbf{Absolute}\\\textbf{error}}
    & \makecell{\textbf{Relative}\\\textbf{error (\%)}}\\
    \midrule
    Hand-held
    & $21$ & $73 \pm 16$ & $8 \pm 13.4$ & $10 \pm 15.5$\\
    Hand-held \& using
    & $21$ & $74 \pm 13.6$ & $9 \pm 14.3$ & $10 \pm 14.2$\\
    \makecell[l]{Trousers\\\ \ back pocket}
    & $2$ & $82 \pm 19$ & $6 \pm 5.7$ & $7 \pm 6.3$\\
    \makecell[l]{Trousers\\\ \ front pocket}
    & $1$ & $75 \pm 0$ & $7 \pm 0.0$ & $10 \pm 0.0$\\
    Handbag
    & $5$ & $94 \pm 9.6$ & $14 \pm 5.5$ & $18 \pm 6.8$\\
    Backpack
    & $16$ & $77 \pm 7.9$ & $5 \pm 3.8$ & $6 \pm 4.7$\\
    Shirt pocket
    & $1$ & $94 \pm 0.0$ & $11 \pm 0.0$ & $13 \pm 0.0$\\
    \bottomrule
  \end{tabular}
\end{table}

\begin{figure}[!h]
  \centering
  \includegraphics[width=.90\linewidth]{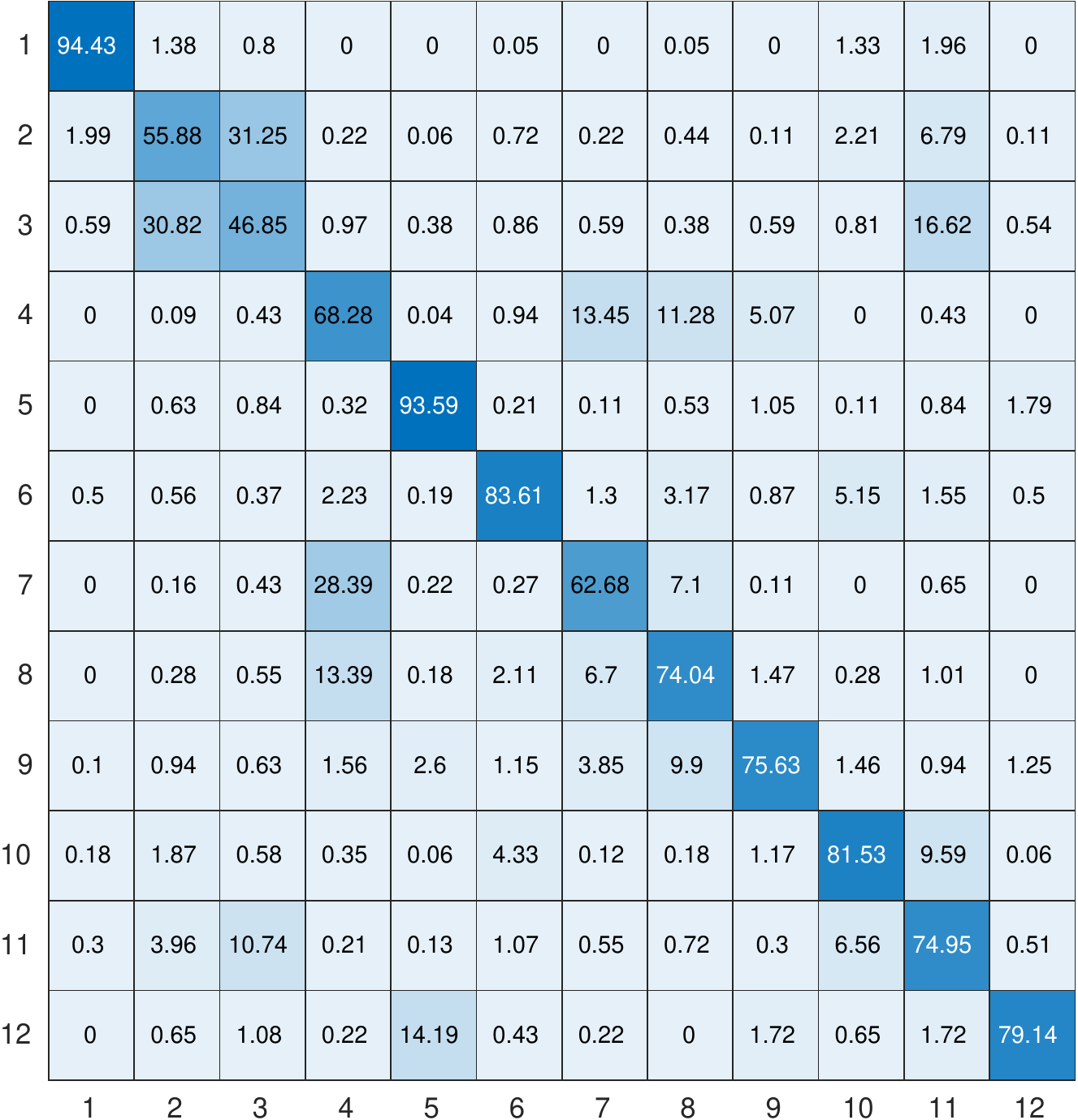}
  \caption{Confusion matrix heat-map for physical activity type recognition of
    \cite{reiss2012creating}. Rows correspond to actual class and columns to
    predicted class. The class labels are 1: lying, 2: sitting, 3: standing, 4:
    walking, 5: running, 6: cycling, 7: Nordic walking, 8: ascending stairs, 9:
    descending stairs, 10: vacuum cleaning, 11: ironing, 12: rope jumping.}
  \label{fig:physical_activity_type_heatmap}
\end{figure}

Figure \ref{fig:physical_activity_type_heatmap} presents a heat-map of the
confusion matrix for the task of physical activity type recognition, based on
the algorithm of \cite{reiss2012creating}. The SVM classifiers have been trained
in a typical leave-one-subject-out (LOSO) fashion. In general, the number of
misclassifications is relatively low, and the actual misclassifications occur
between similar classes. For example, looking at the second and third row and
column of the matrix, significant misclassification is observed. However, these
two classes correspond to sitting and standing, which have very similar
footprints on the mobile phone's accelerometer.

In Table \ref{tab:visited_locations} we present evaluation results for the
algorithm of \cite{luo2017improved} for detecting visited locations based on
location data of \cite{wang2018summary}. We present as TP the number of visited
locations that have been correctly identified by the algorithm using a distance
threshold of $10$ m. We also count the number of visited locations that the
algorithm failed to detect as FN, and the number of erroneous detections as
FP. The algorithm achieves an overall recall rate of more than $0.8$ with a
precision rate of $0.9$.

\begin{table}
  \centering
  \caption{Classification results for visited locations detection using the
    algorithm of \cite{luo2017improved} on the $3$ subjects of the SHL dataset
    \cite{wang2018summary}. Results are shown individually per subject, and
    across all subjects (aggregated as a single subject, not averaged).}
  \label{tab:visited_locations}
  \begin{tabular}{lcccccc}
    \toprule
    \textbf{Subject} & \textbf{TP} & \textbf{FP} & \textbf{FN}
    & \textbf{Precision} & \textbf{Recall} & \textbf{F1-score}\\
    \midrule
    $1$ & $8$ & $0$ & $1$ & $1.00$ & $0.89$ & $0.94$\\
    $2$ & $15$ & $2$ & $4$ & $0.88$ & $0.79$ & $0.83$\\
    $3$ & $13$ & $2$ & $3$ & $0.87$ & $0.81$ & $0.84$\\
    sum & $36$ & $4$ & $8$ & $0.90$ & $0.82$ & $0.86$\\
    \bottomrule
  \end{tabular}
\end{table}

\begin{figure}
  \centering
  \includegraphics[width=.75\linewidth]{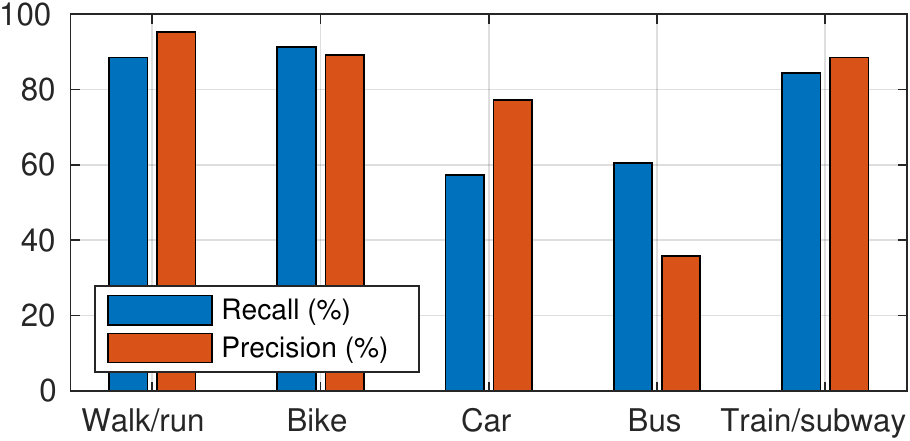}
  \caption{Classification results for transportation mode recognition using the
    algorithm of \cite{shafique2016travel} on the $3$ subjects of the SHL
    dataset \cite{wang2018summary}. Note that ``vehicle'' vs. ``non-vehicle''
    yields recall of $95.41\%$ and precision of $99.44\%$.}
  \label{fig:transportation_mode}
\end{figure}

Finally, Figure \ref{fig:transportation_mode} presents the per-class recall and
precision for transportation-mode detection using the algorithm of
\cite{shafique2016travel}. Both are higher than $0.8$ for walk/run, bike, and
train/subway, and in some cases quite higher (precision for walk/run is over
$0.9$). We have observed that most misclassifications (approximately $39\%$ of
them) are between the car and bus classes, and between the bus and train/subway
classes (approximately $35\%$). This can be attributed to the similarities
between the ``movement'' of these transportation methods. A very encouraging
result, however, is the accuracy between vehicle vs non-vehicle which is $0.98$
(recall and precision are $0.95$ and $0.99$ respectively), since it has a
significant impact in overall physical activity level.

\section{Examples of data views}

The presented system has been integrated in the BigO project and is actively
used. So far, more than $3,000$ students have contributed data by using an
application for Android smart phones and watches. No directly identifiable
information is collected (such as names and e-mail addresses). Students
participate voluntarily (opt-in) with written consent from their parents or
legal guardians where necessary. Ethical approvals have been received for all
pilot studies of the project.

As an example of the different aggregations and data views our system can
provide, we present two examples with data from users of one of the pilot sites,
the Biomedical Research Foundation of the Academy of Athens (BRFAA). Table
\ref{tab:profiles} shows aggregate ``profiles'' for two users with very
different behavior. More detailed information, as well as different types of it,
can be extracted by our proposed system. Collecting such profiles on the large
scale can be used to develop explanatory and predictive models related to
obesity \cite{diou2019methodology}. Table \ref{tab:after_school} groups $72$
users of BRFAA based on their most frequent destination after school. We also
show the average (and std) BMI per group. Such information may be related to
obesogenic behavior and can be objectively, automatically, and unobtrusively
collected by our system.

\begin{table}
  \centering
  \caption{Example profiles of two users of the system.}
  \label{tab:profiles}
  \begin{tabular}{lcc}
    \toprule
    & \textbf{user 1} & \textbf{user 2}\\
    \midrule
    {Gender} & male & female\\
    {Age} & $17$ & $9$\\
    {BMI} & $33$ & $26$\\
    {BMI z-score} & $6$ & $4$\\
    {Average hours of daily monitoring} & $14.16$ & $22.32$\\
    {Daily steps} & $5,891$ & $3,389$\\
    {Daily average activity counts per minute} & $604$ & $216$\\
    {Weekly visits to cafes} & $3.9$ & $1.4$\\
    {Weekly visits to food retailers} & $2.3$ & $1.4$\\
    \bottomrule
  \end{tabular}
\end{table}

\begin{table}
  \centering
  \caption{Most frequent destination after school.}
  \label{tab:after_school}
  \begin{tabular}{lcc}
    \toprule
    \textbf{Destination} & \textbf{Number of users} & \textbf{BMI}\\
    \midrule
    athletics/sports/recreational & $2$ & $30.5 \pm 4.3$\\
    cafe & $5$ & $27.9 \pm 5.1$\\
    fast-food/take-away/restaurant & $12$ & $23.4 \pm 2.1$\\
    home & $24$ & $26.3 \pm 7.9$\\
    unknown & $29$ & $23.0 \pm 4.1$\\
    \bottomrule
  \end{tabular}
\end{table}

\section{Conclusions \& future work}
\label{sec:conclusions}

In this work we have presented a system for collecting big behavioral data from
individuals regarding their physical activity, types of places they visit, and
how they move (transportation). This kind of data can enable the creation of
population-level behavioral profiles per region/neighborhood in order to study
the relationship of such behavior with obesity and the risk for developing
it. To this end, we select, implement, and evaluate algorithms from the
literature that measure physical activity level (number of steps, type of
physical activity), that detect visited places based on location data which can
then be cross-referenced against publicly available map sources, and the
transportation mode used to move from one visited place to another. We evaluate
these algorithms on publicly available datasets and present the results,
demonstrating the feasibility of our proposed system.

\bibliographystyle{IEEEtran}


\end{document}